\documentclass[preprint,10pt,twocolumn]{elsarticle}

\usepackage{epsfig}

\usepackage{amssymb}

\newcommand{\bea}{\begin{eqnarray}}

\newcommand{\eea}{\end{eqnarray}}

\newcommand{\bean}{\begin{eqnarray*}}

\newcommand{\eean}{\end{eqnarray*}}

\newcommand{\nn}{\nonumber \\}

\newcommand{\be}{\begin{equation}}
\newcommand{\ee}{\end{equation}}

\def\spa#1.#2{\langle#1\,#2\rangle}
\def\spb#1.#2{[#1\,#2]}
\def\spab#1.#2.#3{\langle\mskip-1mu{#1} 
                  | #2 | {#3}]}

\def\spba#1.#2.#3{[\mskip-1mu{#1} 
                  | #2 | {#3}\rangle}

\def\spbb#1.#2.#3.#4{[\mskip-1mu{#1} 
                     | {#2} \ {#3} | {#4}]}

\def\spaa#1.#2.#3.#4{\langle\mskip-1mu{#1} 
                     | {#2} \ {#3} | {#4}\rangle}

\def\dea{\langle \ell \ d \ell \rangle}
\def\deb{[d \ell \ \ell]}

\def\nn{{\nonumber}}

\newbox\SlashedBox
\def\slashed#1{\setbox\SlashedBox=\hbox{#1}
\hbox to 0pt{\hbox to 1\wd\SlashedBox{\hfil/\hfil}\hss}#1}
\def\hboxtosizeof#1#2{\setbox\SlashedBox=\hbox{#1}
\hbox to 1\wd\SlashedBox{#2}}

\newbox\charbox
\newbox\slabox
\def\s#1{{      
        \setbox\charbox=\hbox{$#1$}
        \setbox\slabox=\hbox{$/$}
        \dimen\charbox=\ht\slabox
        \advance\dimen\charbox by -\dp\slabox
        \advance\dimen\charbox by -\ht\charbox
        \advance\dimen\charbox by \dp\charbox
        \divide\dimen\charbox by 2
        \raise-\dimen\charbox\hbox to \wd\charbox{\hss/\hss}
        \llap{$#1$}
}}

\begin{document}
\begin{titlepage}

\title{Double-Cut of Scattering Amplitudes and Stokes' Theorem}

\author{ Pierpaolo Mastrolia \\ 
   {\it Theory Group, Physics Department, CERN, 
        CH-1211 Geneva 23, Switzerland}
}

\begin{abstract}

We show how Stokes' Theorem,
in the fashion of the Generalised Cauchy Formula,
can be applied for computing double-cut 
integrals of one-loop amplitudes analytically.
It implies the evaluation of phase-space integrals 
of rational functions in two complex-conjugated variables,
which are simply computed by an indefinite integration in a single variable, 
followed by Cauchy's Residue integration in the conjugated one.
The method is suitable for the cut-construction of the coefficients of 
2-point functions entering the decomposition
of one-loop amplitudes in terms of scalar master integrals.
 
\end{abstract}

\maketitle

\thispagestyle{empty}
\end{titlepage}

Unitarity and analyticity are well-known properties of scattering amplitudes
\cite{OldUnitarity}.
Analyticity grants that amplitudes are determined by 
their own singularity-structure, 
while unitarity grants that the residues at the singular
points factorize into products of simpler amplitudes.
Unitarity and analyticity become tools for the quantitative determinaton of one-loop amplitudes \cite{Bern:1994zx}
when merged with the existence of an underlying representation of amplitudes
as a combination of basic scalar one-loop functions \cite{Passarino:1978jh}.
These functions, known as Master Integrals (MI's),
are $n$-point one-loop integrals, $I_n$ ($1\le n \le 4$),
with trivial numerator, equal to 1, characterised 
by external momenta and internal masses present in the denominator.  
Important improvements of unitarity-based numerical algorithms also make 
use of 
the general structure of one-loop integrands   
\cite{Ossola:2006us,Ellis:2007br,Giele:2008ve,Berger:2008sj}.
In the context of unitarity-based algorithms, 
the issue of computing one-loop amplitudes can be addressed in two stages:
the computation of the coefficients;
and the actual evaluation of the MI's themeselves.
The principle of a unitarity-based method is the
extraction of the coefficients multiplying each MI by
matching the multiparticle cuts of the amplitude onto
the corresponding cuts of the MI's. 

Cutting a propagating particle in an amplitude
amounts to applying the on-shell condition
and replacing its Feynman propagator by the corresponding 
$\delta$-function,
$(p^2 - m^2 + i0)^{-1} \to (2 \pi i) \ \delta^{(+)}(p^2 - m^2) .$
As a result, the original function is substituted by a simpler one,
easier to compute, which, nevertheless, still carries non-trivial
information. 
In fact, the $n$-particle cut of $I_n$ appears in   
the 0-trascendentality term (rational or irrational) of the corresponding 
cut-amplitude, multiplied by the same coefficient 
of $I_n$ in the decomposition of the complete amplitude.
Higher-transcendentality terms, such as logarithms, are associated
to the cuts of higher-point MI's. 

In general, the fulfillment of multiple-cut conditions requires
loop momenta with complex components.
Since the loop momentum has four components, 
the effect of the cut-conditions is to
fix some of them according to the
number of the cuts. Any {\it quadruple}-cut \cite{Britto:2004nc}
fixes the loop-momentum completly, yielding the
algebraic determination of the coefficients 
of $I_n, (n\ge4)$;
the coefficient of 3-point functions, $I_3$,
are extracted from {\it triple}-cut 
\cite{MastroliaTriple,FordeTriBub,BjerrumBohr:2007vu,Kilgore:2007qr,Badger:2008cm};
the evaluation of {\it double}-cut 
\cite{Britto:2005ha,Britto:2006sj,ABFKM,FordeTriBub,Britto:2007tt,Kilgore:2007qr,Britto:2008vq,Britto:2008sw,Badger:2008cm}
is necessary for extracting the coefficient of 2-point
functions, $I_2$; and 
finally, in processes involving massive particles,
the coefficients of 1-point functions, $I_1$, are detected by 
{\it single}-cut \cite{Kilgore:2007qr,NigelGlover:2008ur,Britto:2009wz}.
In cases where fewer than four denominators are cut, the loop momentum
is not frozen: the free-components are left over as phase-space 
integration variables.

In this letter, we show a novel efficient method for the 
analytic evaluation of the coefficients of one-loop 2-point 
functions {\it via} double-cuts.
Spun-off from the spinor-integration technique 
\cite{Britto:2005ha,Britto:2006sj,ABFKM,Britto:2007tt,Britto:2008vq,Britto:2008sw},
the method hereby presented is an 
application of Stokes' Theorem.
We analyze the double-cut of massless particles in
four-dimensions, which also is the essential ingredient
for the phase-space integration in the general case
of one-loop massive amplitudes in dimensional-regularization
\cite{ABFKM,Britto:2007tt,Britto:2008vq,Britto:2008sw,MastroliaTriple}. \\
Due to a special decomposition of the loop-momentum,
the double-cut phase-space integral is written
as parametric integration of rational function 
in two complex-conjugated variables.
By applying Stokes' Theorem, the integration is carried on in 
two simple steps:
an indefinite integration in one variable, followed by Cauchy's
Residue Theorem in the conjugated one. 

The coefficients of the 2-point scalar functions, being proportional 
to the rational term of the double-cut, can be directly 
extracted from the indefinite integration by Hermite Polynomial Reduction. 

In a framework where factorization properties of scattering
amplitudes are accessed {\it via} complex momenta, 
the double-cut integration presented here can be considered
as the natural extension of the technique used 
to prove BCFW-recurrence relation for tree-level
amplitudes \cite{BCFWproof}. 
In the latter case, scattering tree-amplitudes are
holomorphic functions, depending only on
one complex variable, and Cauchy's Residue Theorem 
is sufficient for their complete determination. In the case 
of the double-cut of one-loop amplitudes, where the integrand depends 
on two complex-cojugated variables, Stokes' Theorem
in the fashion of Generalised Cauchy Formula, becomes the driving principle.

\section{Double-Cut}
\label{sec:integration}

\noindent
{\it -- Phase-Space Parametrization. }
The starting point of our derivation
is the spinorial parametrization of
the Lorentz invariant phase-space (LIPS) in the $K^2$-channel
\cite{Cachazo:2004kj,Britto:2005ha,Britto:2006sj},
\bea
\int d^4\Phi \!\!\!\!&\equiv&\!\!\!\!\! 
\int d^4 \ell_1 
\ \delta^{(+)}(\ell_1^2) 
\ \delta^{(+)}((\ell_1-K)^2) 
= \nn \\
\!\!\!\!&=&\!\!\!\!\!
 \int \!\!\!\! \int
{\dea \deb  \over \spab\ell.K.\ell} \!\!
\int \!\! t dt \ 
 \delta^{(+)}\bigg(t - { K^2 \over \spab\ell.K.\ell} \bigg) , \quad 
\label{def:phi4}
\eea
obtained by rescaling the original
loop-variable $\ell_1^\mu$ as,
\bea  
\ell_1^\mu = {\spab{\ell_1}.\gamma^\mu.{\ell_1} \over 2} 
\equiv 
t \ \ell^\mu 
= 
t {\spab{\ell}.\gamma^\mu.\ell \over 2} \ , 
\eea
with $\ell_1^2=\ell^2=0$.
In terms of spinor variables, the rescaling reads,
\bea
|\ell_1\rangle = \sqrt{t} \ |\ell\rangle \ , \qquad
|\ell_1] = \sqrt{t} \  |\ell] \ ,
\label{def:rescaling}
\eea
where $t$, the rescaling parameter, is frozen
as a consequence of the (second of the) on-shell conditions,
and $\ell^\mu$ becomes the new loop integration variable.

\noindent
{\it -- Change of Variables. }
We take two massless momenta,
say $p_\mu$ and $q_\mu$ 
fulfilling the conditions,
\bea
&& \hspace*{-0.5cm}
p_\mu + q_\mu = K_\mu \ , \nn \\
&& \hspace*{-0.5cm}
p^2=q^2=0 \ , \quad 2 p\cdot q = 2 p\cdot K = 2 q\cdot K \equiv K^2 \ , \qquad
\label{def:specialpq}
\eea
and decompose $\ell_\mu$ in a basis
of four massless momenta constructed out of them,
\bea
\ell_\mu =
p_\mu + 
z \ \bar{z}\ q_\mu
+  {z \over 2} \ { \spab q.\gamma_\mu.p }   
+  {\bar{z} \over 2} \ { \spab p.\gamma_\mu.q } \ .
\label{def:loopdeco}
\eea
Notice that the vectors ${ \spab q.\gamma_\mu.p \over 2}$ and 
${ \spab p.\gamma_\mu.q \over 2}$ are trivially orthogonal to both 
$p_\mu$ and $q_\mu$. 
The above decomposition can be realized
starting from the definition of $\ell_\mu$
in terms of spinor variables,
$
\ell^\mu \ 
=  
{\spab{\ell}.\gamma^\mu.\ell \over 2} \ ,
$
and performing the following spinor decomposition,
\bea
|\ell\rangle \equiv |p\rangle + z |q\rangle \ , \quad
|\ell] \equiv |p] + \bar{z} |q]  \ .
\label{def:spinordeco}
\eea
By changing variables $(|\ell\rangle, |\ell]) \to (z, \bar{z})$ 
as in (\ref{def:spinordeco}),
and using (\ref{def:specialpq}), one can write,
\bea
\dea \deb \!\!\!\!&=&\!\!\!\! K^2 \ dz \ d\bar{z} \ , \quad \\
\spab \ell.K.\ell \!\!\!\!&=&\!\!\!\! K^2 \ (1 + z \bar{z}) \ .
\eea
Hence, the LIPS in (\ref{def:phi4}) reduces to the novel form,
\bea
\int d^4 \Phi \!\!\!\!&=&\!\!\!\! 
\oint
d z \int  d\bar{z} 
\int dt \ t^2  \  
\delta \bigg(t - { 1 \over (1 +  z \bar{z})} \bigg) \ , \quad
\label{eq:novelphi4}
\eea
where $t$ is a positive quantity as assured by the argument
of the $\delta$-function.

\begin{figure}[t]
\begin{center}
\vspace*{-1cm}
\includegraphics[scale=0.15]{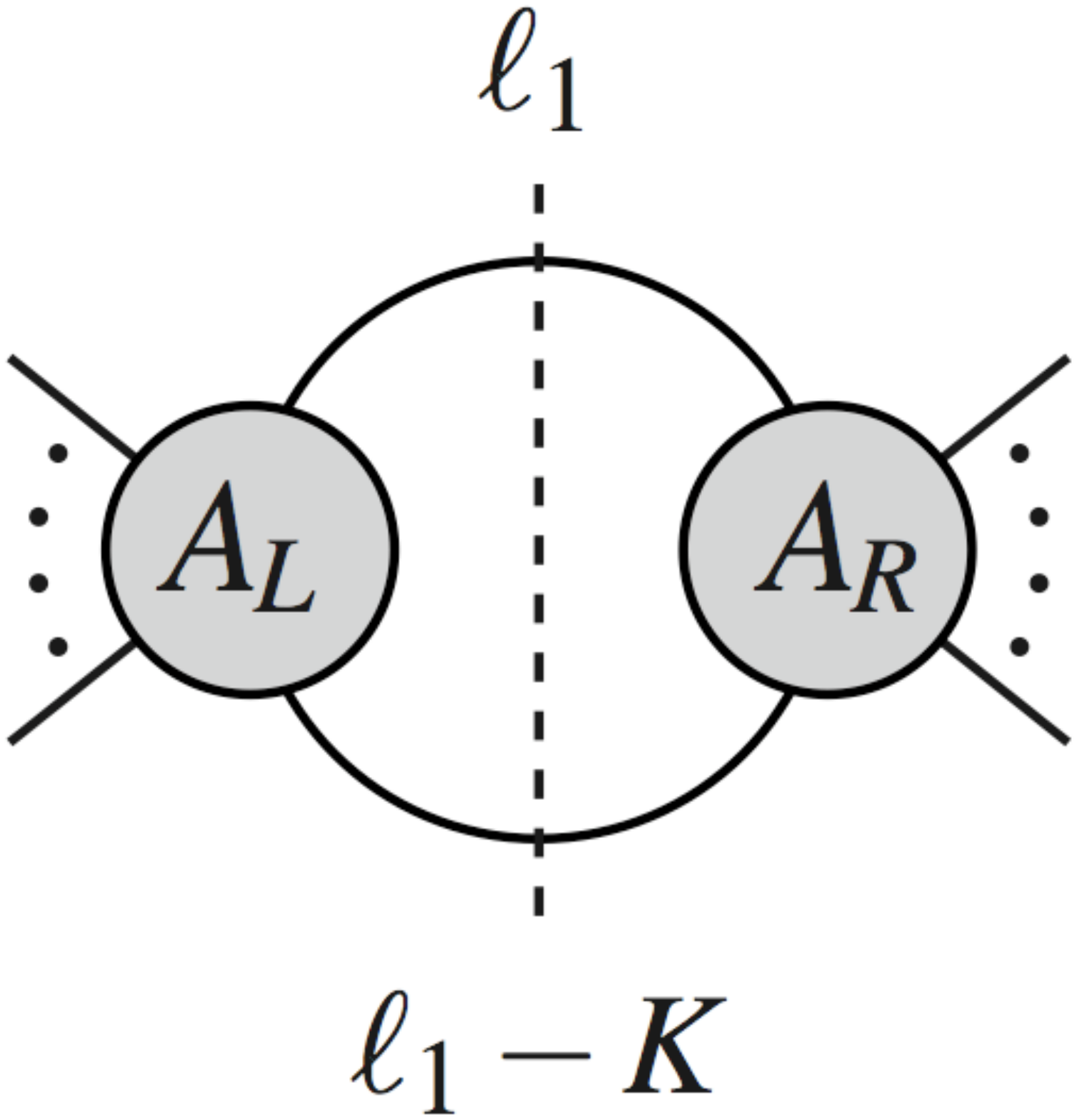}
\vspace*{-1.0cm}
\caption{Double-cut of one-loop amplitude in the $K^2$-channel.}
\label{fig:doublecut}
\end{center}
\end{figure}

\noindent
{\it -- Double-Cut Integration. }
The double-cut of a generic $n$-point amplitude
in the $K^2$-channel is defined as
\bea
\Delta \equiv 
\int d^4\Phi \  
A^{\rm tree}_L(|\ell_1\rangle, |\ell_1]) \ 
A^{\rm tree}_R(|\ell_1\rangle, |\ell_1])  \ , \ 
\eea
where $A^{\rm tree}_{L,R}$ are the tree-level amplitudes
sitting at the two sides of the cut, see Fig.\ref{fig:doublecut}.
After rescaling $\ell_1^\mu$ as in (\ref{def:rescaling}),
and using expression (\ref{eq:novelphi4}) for the LIPS, one has,
\bea
\Delta \!\!\!\!&=&\!\!\!\!
\int d^4\Phi \  
A^{\rm tree}_L(t, |\ell\rangle, |\ell]) \ 
A^{\rm tree}_R(t, |\ell\rangle, |\ell]) \ \nn \\
\!\!\!\!&=&\!\!\!\!
 \oint
d z \int  d\bar{z} 
\int {t^2 \ dt } \  
 \delta\bigg(t - { 1 \over (1 +  z \bar{z})} \bigg) \ 
\times \nn \\
\!\!\!\!& &\!\!\!\! \quad
 t^{\alpha_L + \alpha_R} \ 
A^{\rm tree}_L(|\ell\rangle, |\ell]) \ 
A^{\rm tree}_R(|\ell\rangle, |\ell]) \ ,
\label{eq:Deltaalphas}
\eea
whera $\alpha_{L,R}$ parametrizes the scaling behaviour of $A^{\rm tree}_{L,R}$.
The $t$-integration can be performed trivially, because of 
the presence of the $\delta$-function. 
Then, by using the decomposition (\ref{def:loopdeco}, \ref{def:spinordeco}),
the double-cut becomes a double-integral, 
\bea
\Delta =
 \oint
d z \int  d\bar{z} \ f(z,\bar{z}) \ ,
\label{dummydef:zbarzInt}
\eea
where $f$ is a rational function of $z$ and $\bar{z}$.
As such, it can be expressed as a ratio of two polynomials,
say $P$ and $Q$,
\bea
f(z,\bar{z}) 
=
{A^{\rm tree}_L(z,\bar{z}) \ A^{\rm tree}_R(z,\bar{z}) 
\over (1 + z\bar{z})^{\alpha_L+\alpha_S+1} } 
=  {P(z,\bar{z}) \over Q(z,\bar{z})} \ ,
\label{eq:effdef}
\eea
with the following relations between their degrees,
\bea
 {\rm deg}_z Q = {\rm deg}_z P + 2  \ , 
\qquad
 {\rm deg}_{\bar{z}} Q = {\rm deg}_{\bar{z}} P + 2  \ .
\label{eq:degrees}
\eea
We remark that the {\it double integration} 
in $z$- and $\bar{z}$-variables
appearing in Eq.(\ref{dummydef:zbarzInt}) will be properly 
justified in Sec.\ref{sec:theproof}.
For the moment, with abuse of notation, we simlpy denote it as 
a convolution of an indefinite $\bar{z}$-integral 
and a contour $z$-integral, which are the actual operations
we are going to carry out. \\
To begin with the integration, 
we find a primitive of $f$ with respect
to $\bar{z}$, say $F$, by keeping $z$ as independent variable,
\bea
F(z,\bar{z}) 
\!\!\!\!&=&\!\!\!\! \int d\bar{z} \ f(z,\bar{z})  \ , 
\label{eq:primitive}
\eea
so that $\Delta$ becomes,
\bea
\Delta = 
\oint dz \ F(z, \bar{z})  
= \oint dz \int d\bar{z} \ F_{\bar{z}} \ ,
\label{eq:dummyDelta}
\eea
where $F_{\bar{z}}$ is a short-hand notation for
${\partial F / \partial{\bar{z}}}$.
Before proceeding with the final integration on the $z$-variable,
let us analyse the structure of $F$.
Since $F$ is the primitive of a rational function, its general
form can only contain two types of terms: 
a rational term and a logarithimc one,
\bea
F(z,\bar{z}) = F^{\rm rat}(z,\bar{z}) + F^{\rm log}(z,\bar{z}) \ .
\label{eq:EFFdef}
\eea
It is important to notice that the presence of the term
$F^{\rm rat}$ depends on the powers of $t$ in Eq.(\ref{eq:Deltaalphas}):
$F^{\rm rat}$ can be generated, after integrating $f$ in $\bar{z}$, only
if $\alpha_R + \alpha_L \ge 0$. 
The $z$-integration will be performed by applying Cauchy's Residue Theorem,
therefore the final structure of the double-cut is determined by the 
nature of $F$. Namely, the $z$-integration of $F^{\rm rat}$ [$F^{\rm log}$]
is responsible of the rational [logarithmic] term
of $\Delta$. 
 
\noindent
{\it -- 2-point Function. }
We also know {\it apriori} that the double-cut of a 2-point scalar function
in 4-dimension is a rational, or better (to account
for the massive case as well) a non-logarithmic term; 
while the double-cut of higher-point scalar functions might contain 
logarithms (with $K^2$-dependent argument).
Hence, the coefficient of a 2-point function in the $K^2$-channel
will appear in $\Delta^{\rm rat}$, the integration of $F^{\rm rat}$ in $z$,
\bea
\Delta^{\rm rat} \equiv \oint dz \ F^{\rm rat}(z, \bar{z}) \ ,
\eea
where the $z$-integration is performed {\it via} 
Cauchy's Residue Theorem. 
The integrand $F^{\rm rat}$ is rational in $z$,
and contains poles whose location in the complex 
plane is a unique signature of the Feynman integral they come from
 \cite{Britto:2007tt,Britto:2008vq,ArkaniHamed:2008gz}.
The choice of $p$ and $q$ specified in Eqs.(\ref{def:specialpq})
grants that there exists a pole at $z=0$ associated to the 
2-point function in the $K^2$-channel, $I_2(K^2)$;
while the reduction of higher-point functions that have $I_2(K^2)$ 
as subdiagram can generate poles at finite $z$-values.  
Because of the presence of $\bar{z}$, 
through the term $(1+ z \bar{z})$, $F^{\rm rat}$ is non-analytic.
The Residue Theorem has to be applied by reading the residues
in $z$, and substituting the corresponding complex-conjugate values
where $\bar{z}$ appears.
Therefore, the result of $\Delta^{\rm rat}$ can be implicitly written as,
\bea
\Delta^{\rm rat} \!\!\!\!&=& \!\!\!\! 2 \pi i \Big(
{\rm Res}_{z = 0}F^{\rm rat}(z, \bar{z}) + 
{\rm Res}_{z \ne 0}F^{\rm rat}(z, \bar{z}) \Big)\ . \quad \nn \\
\label{eq:DeltaRatfinal}
\eea

\noindent
{\it -- Double-cut of the Scalar Function $I_2$. }
Let us evaluate the double-cut of the 2-point scalar 
function $I_2$, which also is a prototype example:
\bea
\Delta I_2 
\!\!\!\!&=& \!\!\!\!
\int d^4\Phi = \nn \\
\!\!\!\!&=& \!\!\!\!
 \oint
d z \int  d\bar{z} 
\int {t^2 \ dt } \  
 \delta\bigg(t - { 1 \over (1 +  z \bar{z})} \bigg) 
= \nn \\
\!\!\!\!&=& \!\!\!\!
 \oint
d z \int  d\bar{z} \ 
 { 1 \over (1 +  z \bar{z})^2} = \nn \\
\!\!\!\!&=& \!\!\!\!
 \oint
d z  \ 
 { (-1) \over (1 +  z \bar{z}) \ z} \ ,
\eea
where we notice that the primitive in $\bar{z}$,
called $F$ in (\ref{eq:primitive}),
has only a rational term (the logarithmic contribution is absent),
\bea
F(z,\bar{z}) = \!\! \int \!\! d\bar{z} \ f(z,\bar{z}) = 
{ (-1) \over (1 +  z \bar{z}) \ z} \equiv  F^{\rm rat}(z,\bar{z}) .
\label{eq:EFF4I2}
\eea
For the last integration in $z$, by applying the Residue Theorem,
we take the residue of the unique simple pole at $z=0$, since
the term $(1 +  z \bar{z}) = (1 +  |z|^2)$  never
vanishes, being always positive. 
The final result of the double-cut of the scalar 2-point
function reads,
\bea
\Delta I_2 
\!\!\!\!&=& \!\!\!\!
(2 \pi i) \ {\rm Res}_{z = 0}F^{\rm rat}(z, \bar{z}) = - 2 \pi i \ .
\label{eq:DeltaI2final}
\eea

\noindent
{\it -- Coefficient of the 2-point Function. }
The expression of the 2-point coefficient can be finally 
obtained by taking the ratio of $\Delta^{\rm rat}$ in (\ref{eq:DeltaRatfinal}) 
and the double-cut of $I_2$ in (\ref{eq:DeltaI2final}),
\bea
c_2 \!\!\!\!&\equiv&\!\!\!\! 
  {\Delta^{\rm rat} \over \Delta I_2} = \nn \\
    \!\!\!\!&=&\!\!\!\!
 - {\rm Res}_{z = 0}F^{\rm rat}(z, \bar{z}) - 
{\rm Res}_{z \ne 0}F^{\rm rat}(z, \bar{z}) \ . \quad
\label{eq:c2final}
\eea

\noindent
{\it -- Hermite Polynomial Reduction. }
To optimize the integration algorithm, one can use 
the so called Hermite Polynomial Reduction (HPR), 
a technique enabling the direct extraction
of the rational term of the primitive of a rational function,
without computing the integral as a whole.
Based on the square-free factorization of the integrand, 
HPR can be used to write the result
of any integral of a rational
function as a pure rational term plus
another integral that, if explicitly computed, would generate
the logarithmic remainder.

As written in Eq.(\ref{eq:c2final}), the coefficient
of the 2-point function comes only from the term $F^{\rm rat}$, and not 
from $F^{\rm log}$, see Eqs.(\ref{eq:primitive}, \ref{eq:EFFdef});
$F^{\rm rat}$ is the rational term in the result of the
$\bar{z}$-integration of $f$, which is rational in $\bar{z}$, 
see Eq.(\ref{eq:effdef}).
Therefore HPR is suitable for extracting $F^{\rm rat}$ out
of the $\bar{z}$-integration.

The integration algorithm of Sec.\ref{sec:integration} 
can be implemented with {\tt S@M} \cite{Maitre:2007jq}
together with the routine \cite{HPR} for Hermite Polynomial Reduction.

\section{Stokes'  Theorem}
\label{sec:theproof}

In this section we give a formal definition
of the $z$-$\bar{z}$ integration used in Sec.\ref{sec:integration},
as an application of Stokes' Theorem for differential forms. 
In what follows, we use the notation:  
$g_z = \partial g/\partial z$ 
and $g_{\bar{z}} = \partial g/\partial \bar{z}$.

Let us recall that the complex 1-form 
\bea
\chi = {1 \over z-z_0} dz \ ,
\eea 
which is defined for all $z$ except $z_0$, is a closed form,
\bea
d\chi = d\bigg( {1 \over z-z_0}\bigg) \wedge dz 
      =  {(-1) \over (z-z_0)^{2}} dz \wedge dz = 0 \ .
\eea
We consider any complex smooth function ${\cal F}$ and differentiate 
the 1-form $\omega = \cal{F}\chi$, 
\bea 
\omega = (z-z_0)^{-1} {\cal F} dz \ ,
\label{eq:omega}
\eea
obtaining the 2-form,
\bea
d\omega =  
d{\cal F} \wedge \chi = 
     (z-z_0)^{-1} {\cal F}_{\bar{z}} \ d\bar{z} \wedge dz \ .
\label{eq:deomega}
\eea
Now we take a domain $D$ in the complex plane and 
apply Stokes' Theorem to $d\omega$.
Due to the singularity of $\omega$ at $z_0$, we remove a tiny disk 
$D(z_0;r)$, centered at $z_0$ with radius $r$, from $D$.
Then $\omega$ has no singularity in the regulated domain $D_r = D - D(z_0;r)$,
and we may apply Stokes' Theorem:
\bea
\int \!\!\!\! \int_{D_r} d\omega = 
\int_{\partial D_r} \omega
= \int_{\partial D} \omega - \int_{\partial D(z_0;r)} \omega \ .
\label{eq:stokes1}
\eea
Here $\partial D(z_0;r)$ is a circle $\gamma$ around the point $z_0$,
which is described by the parametric equation $\gamma(t) = z_0+re^{it}$.
Since ${\cal F}(z_0+re^{it})$ converges to ${\cal F}(z_0)$ as the radius $r$ 
shrinks to 0, the last integral in Eq.(\ref{eq:stokes1}),
\bea
\int_{\partial D(z_0;r)} \omega 
\!\!\! &=& \!\!\!
i \int_{0}^{2\pi} {\cal F}(z_0+re^{it}) dt \ ,
\eea
converges to $2 \pi i {\cal F}(z_0)$ as $r$ goes to 0. 
Letting $r \to 0$ in Eq.(\ref{eq:stokes1}), the disk 
$D(z_0;r)$ disappears and $D_r$ fills up $D$.
Consequently Stokes' Theorem can be reformulated as,
\bea
\int \!\!\!\! \int_{D} d\omega
= \int_{\partial D} \omega - 2 \pi i {\cal F}(z_0)  \ .
\label{eq:stokes}
\eea
By using the explicit expression of $\omega$ 
and $d\omega$, in Eqs.(\ref{eq:omega}, \ref{eq:deomega}),
and rearranging terms, we obtain
the so called {\it Generalised Cauchy Formula} 
or {\it Cauchy-Pompeiu Formula},
\bea
2 \pi i {\cal F}(z_0) =
 \int_{\partial D} { {\cal F}(z) \over z-z_0} dz
- \int \! \! \! \! \int_D { {\cal F}_{\bar{z}} \over z-z_0} d\bar{z} \wedge dz.
\label{def:GenCauchy}
\eea
Let us discuss two special cases.\\
First, when ${\cal F}$ is analytic, ${\cal F}_{\bar{z}} = 0$, hence 
we obtain,
\bea
 {\cal F}(z_0) = {1 \over 2 \pi i}
 \int_{\partial D} { {\cal F}(z) \over z-z_0} dz
\eea
which is the well-known Cauchy Formula,
where ${\partial D}$ is any closed curve surrounding $z_0$. \\
Secondly, when ${\cal F}$ vanishes on the boundary of $D$,
that is ${\cal F}|_{\partial D} = 0$, Eq.(\ref{def:GenCauchy}) becomes,
\bea
 {\cal F}(z_0) = {1 \over 2 \pi i}
\int \! \! \! \! \int_D { {\cal F}_{\bar{z}} \over z-z_0} dz \wedge d\bar{z}.
\label{eq:result}
\eea
where we used $d\bar{z} \wedge dz = - dz \wedge d\bar{z}$. \\
The expression (\ref{eq:result}) is what needed 
to define properly the double-cut $\Delta$ given 
in Eqs.(\ref{dummydef:zbarzInt}, \ref{eq:dummyDelta}),
which we rewrite here as,
\bea
\Delta \equiv 
  \int \!\!\!\! \int_D f(z,\bar{z}) \  dz \wedge d\bar{z} 
= \int \!\!\!\! \int_D { {\cal F}_{\bar{z}} \over z-z_0} dz \wedge d\bar{z} \ ,
\eea
by identifying $
f = F_{\bar{z}} = { {\cal F}_{\bar{z}} / (z-z_0)}$,
and $ 
F = { {\cal F} / (z-z_0)}$, where
the functions $f$ and $F$ were defined in 
Eqs.(\ref{eq:effdef}, \ref{eq:primitive}, \ref{eq:EFFdef}).
The integration domain, $D$,  is the whole complex plane. 
The vanishing of ${\cal F}$ on the boundary is granted
by the structure of the rational integrand and relations
(\ref{eq:degrees})
among the degrees of numerator and denominator. \\
To deal with the general case, where more than one pole might appear,
the calcualtion of $\Delta$ trivially generalises, 
by the superimposition principle, to the sum of the residues
at all the poles in $z$,
\bea 
\Delta 
\!\!\!\!&\equiv&\!\!\!\!
\int \!\!\!\! \int_D F_{\bar{z}} \ dz \wedge d\bar{z} = \nn \\
\!\!\!\!&=&\!\!\!\!
\sum_j
\int \!\!\!\! \int_D { {\cal F}_{\bar{z}}^{(j)} 
\over z-z_j} dz \wedge d\bar{z} \nn \\
\!\!\!\!&=&\!\!\!\!
 2 \pi i \sum_{j \in {\rm poles} } {\cal F}^{(j)}(z_j) \ , \qquad 
\label{eq:finalresult}
\eea
due to the subtraction of a disk around each of the $z$-poles 
from the domain $D$. \\
Finally, Eq.(\ref{eq:finalresult}) 
validates Eq.(\ref{eq:DeltaRatfinal}), hence
the expression for the coefficient $c_2$ in Eq.(\ref{eq:c2final}). 
Notice that the role of $z$ and $\bar{z}$ in the application
of Stokes' Theorem can be interchanged, reflecting the symmetry 
of $c_2$ under the exchange  $p \leftrightarrow q$ in (\ref{def:specialpq}). \\

\noindent
{\it -- Acknowledgements. }
I am indebted to Ed Witten for 
inviting me to consider Stokes' Theorem as the proper
formal framework for the twofold complex integration
hereby presented.
I also thank Mario Argeri, Simon Badger, Michele Caffo,
Bryan Lynn,  Bob McElrath, Stefano Pozzorini,
and Ciaran Williams for stimulating discussions.

\end{document}